\documentclass[11pt, letterpaper]{article}

\usepackage[letterpaper,margin=1in,marginparwidth=0.8in]{geometry}
\usepackage{ifpdf}
\ifx\pdfoutput\undefined
\pdffalse
\else
\pdfoutput=1
\pdftrue
\fi

\ifpdf
\usepackage[pdftex]{graphicx}
\else
\usepackage{graphicx}
\fi

\usepackage[utf8]{inputenc}     
\usepackage[T1]{fontenc}
\usepackage{times}              
\usepackage{amsmath}            
\usepackage{amsfonts}
\usepackage{txfonts}            

\usepackage[numbers,sort&compress]{natbib}
\usepackage{varioref}
\usepackage[usenames]{xcolor}
\usepackage{listings}

\usepackage[hidelinks,bookmarksopen=true,bookmarksopenlevel=2]{hyperref}

\usepackage{booktabs}
\usepackage{longtable}
\usepackage{multirow}
\usepackage[font=it,labelfont=bf]{caption}
\usepackage{rotating}
\usepackage{enumitem}
\usepackage{ifthen}
\usepackage{rotating}
\usepackage{calc}                  
\usepackage{fancyhdr}
\usepackage{caption,subcaption}
\usepackage{wrapfig}
\usepackage{pdfpages}
\usepackage{verbatim}
\usepackage[titletoc]{appendix}

\usepackage[compact]{titlesec}
\newlist{tightlist}{itemize}{1}
\setlist[tightlist]{label=\textbullet,nosep,leftmargin=*,
  before=\vspace{-12pt},after=\vspace{-2pt}}

\newcommand{\smartinclude}[2]{
  \IfFileExists{#2.tex}{
    \input{#2}
  }{
    \IfFileExists{#2.pdf}{

      \markright{#1}
      \phantomsection 
      \addcontentsline{toc}{section}{\numberline {}#1}

      \includepdf[pages=-,pagecommand={\thispagestyle{fancy}}]{#2}
    }{

      \ifthenelse{\boolean{printmissingintoc}}{
        \markright{#1}
        \phantomsection 
        \addcontentsline{toc}{section}{\numberline {}#1}
      }{}

      \ifthenelse{\boolean{printmissingwarnings}}{
        \begin{center} 
          \textbf{\color{red}\Large No file found for '#1'!\\
            Tried #2.tex and #2.pdf} \\
        \end{center}
        \typeout{smartinclude: WARNING: File not found: #2.tex or #2.pdf}
      }{}
    }     
  }
}

\newcommand{\smartincludeempty}[2]{
  \IfFileExists{#2.tex}{
    \input{#2}
  }{
    \IfFileExists{#2.pdf}{

      \markright{#1}
      \phantomsection 
      \addcontentsline{toc}{section}{\numberline {}#1}

      \includepdf[pages=-,pagecommand={\thispagestyle{empty}}]{#2}
    }{

      \ifthenelse{\boolean{printmissingintoc}}{
        \markright{#1}
        \phantomsection 
        \addcontentsline{toc}{section}{\numberline {}#1}
      }{}

      \ifthenelse{\boolean{printmissingwarnings}}{
        \begin{center}
          \textbf{\color{red}\Large No file found for '#1'!\\
            Tried #2.tex and #2.pdf} \\
        \end{center}
        \typeout{smartinclude: WARNING: File not found: #2.tex or #2.pdf}
      }{}
    }     
  }
}

\newcommand{\smartincludeplain}[2]{
  \IfFileExists{#2.tex}{
    \input{#2}
  }{
    \IfFileExists{#2.pdf}{

      \markright{#1}
      \phantomsection 
      \addcontentsline{toc}{section}{\numberline {}#1}

      \includepdf[pages=-,pagecommand={\thispagestyle{plain}}]{#2}
    }{

      \ifthenelse{\boolean{printmissingintoc}}{
        \markright{#1}
        \phantomsection 
        \addcontentsline{toc}{section}{\numberline {}#1}
      }{}

      \ifthenelse{\boolean{printmissingwarnings}}{
        \begin{center}
          \textbf{\color{red}\Large No file found for '#1'!\\
            Tried #2.tex and #2.pdf} \\
        \end{center}
        \typeout{smartinclude: WARNING: File not found: #2.tex or #2.pdf}
      }{}
    }     
  }
}

\newcommand{\smartincludeemptych}[2]{
  \IfFileExists{#2.tex}{
    \input{#2}
  }{
    \IfFileExists{#2.pdf}{

      \markboth{#1}{}
      \phantomsection 
      \addstarredchapter{#1}

      \includepdf[pages=-,pagecommand={\thispagestyle{empty}}]{#2}
    }{

      \ifthenelse{\boolean{printmissingintoc}}{
        \markboth{#1}{}
        \phantomsection 
        \addstarredchapter{#1}
      }{}

      \ifthenelse{\boolean{printmissingwarnings}}{
        \begin{center}
          \textbf{\color{red}\Large No file found for '#1'!\\
            Tried #2.tex and #2.pdf} \\
        \end{center}
        \typeout{smartinclude: WARNING: File not found: #2.tex or #2.pdf}
      }{}
    }     
  }
}

\newcommand{\smartincludeplainch}[2]{
  \IfFileExists{#2.tex}{
    \input{#2}
  }{
    \IfFileExists{#2.pdf}{
      \markboth{#1}{}
      \phantomsection 
      \addstarredchapter{#1}

      \includepdf[pages=-,pagecommand={\thispagestyle{plain}}]{#2}
    }{

      \ifthenelse{\boolean{printmissingintoc}}{
        \markboth{#1}{}
        \phantomsection 
        \addstarredchapter{#1}
      }{}

      \ifthenelse{\boolean{printmissingwarnings}}{
        \begin{center}
          \textbf{\color{red}\Large No file found for '#1'!\\
            Tried #2.tex and #2.pdf} \\
        \end{center}
        \typeout{smartinclude: WARNING: File not found: #2.tex or #2.pdf}
      }{}
    }     
  }
}

\title{Unraveling Diffusion in Fusion Plasma: A Case Study of In Situ Processing and Particle Sorting}

\author{Junmin Gu, Paul Lin, Kesheng Wu,\\
 Lawrence Berkeley National Laboratory, Berkeley, CA, \\ \\
 Seung-Hoe Ku, C.S. Chang, R. Michael Churchill,\\
 Princeton Plasma Physics Laboratory, Princeton, NJ, \\ \\
Jong Choi, Norbert Podhorszki, Scott Klasky\\
Oak Ridge National Laboratory, Oak Ridge, TN}
\date{}

\begin{document}
 
\maketitle

\pagestyle{plain}
\begin{abstract}
This work starts an \textit{in situ} processing capability to study a certain diffusion process in magnetic confinement fusion.
This diffusion process involves plasma particles that are likely to escape confinement.
Such particles carry a significant amount of energy from the burning plasma inside the tokamak to the diverter and damaging the diverter plate.
This study requires \textit{in situ} processing because of the fast changing nature of the particle diffusion process.
However, the \textit{in situ} processing approach is challenging because the amount of data to be retained for the diffusion calculations increases over time, unlike in other \textit{in situ} processing cases where the amount of data to be processed is constant over time.
Here we report our preliminary efforts to control the memory usage while ensuring the necessary analysis tasks are completed in a timely manner.
Compared with an earlier naive attempt to directly computing the same diffusion displacements in the simulation code, this \textit{in situ} version reduces the memory usage from particle information by nearly 60\% and elapsed time by about 20\%.
\end{abstract}

\section{Introduction}
\label{sec:intro}
Understanding the dynamics of the fusion plasma at the edge of magnetic confinement is critical for developing stable controlled fusion energy.
Toward this goal, the X-point Gyrokinetic Code (XGC)~\cite{xgc2008,XGC, xgc2009} 
needs to track particles near the edge, especially those crossing the magnetic separatrix surface; see an illustration in Fig.~\ref{fig:outline}(a).
This work is a preliminary study of the diffusion process of those particles with high likelihood of escaping confinement.
These particles could remove mass and heat from the fusion plasma to the scrape-off layer outside the separatrix surface~\cite{Kube2022diverter, Chang:2015:ELM}.
Since this process of escaping could happen very quickly, it is therefore necessary to track the particle positions at every simulation time step.
If the particle trajectories are stored in memory, it would significantly increase the memory requirements for XGC.
An early attempt to compute statistics representing the diffusion process within the XGC simulation code was found to significantly increase the memory requirement per particle (more detail in Section~3.1). 
This additional memory requirement was considered to be too high for memory-hungry leadership-class computers; therefore, the team started to explore \textit{in situ} processing options.

However, the team encountered a new problem to solve.  In a typical \textit{in situ} processing use case, the memory required on each analysis node is fixed over time, while our diffusion calculations involve an unknown number of particles and a growing number of simulation time steps.
This short paper describes our exploration of implementing this \textit{in situ} processing for diffusion calculation with ADIOS~\cite{softwareX}, with the main focus on (1) how to select the particles to limit the number of particles to be tracked, (2) how to sort the particles to build the particle trajectories with a minimal memory requirement, and (3) how to select the specific quantities to compute to represent the diffusion processes.

\section{Background}
\label{sec:background}
In this section we will briefly describe the foundation of our work: the XGC application and the I/O library ADIOS.

\begin{figure}
\centering
\begin{tabular}[b]{cc}
\includegraphics[width=0.45\textwidth]{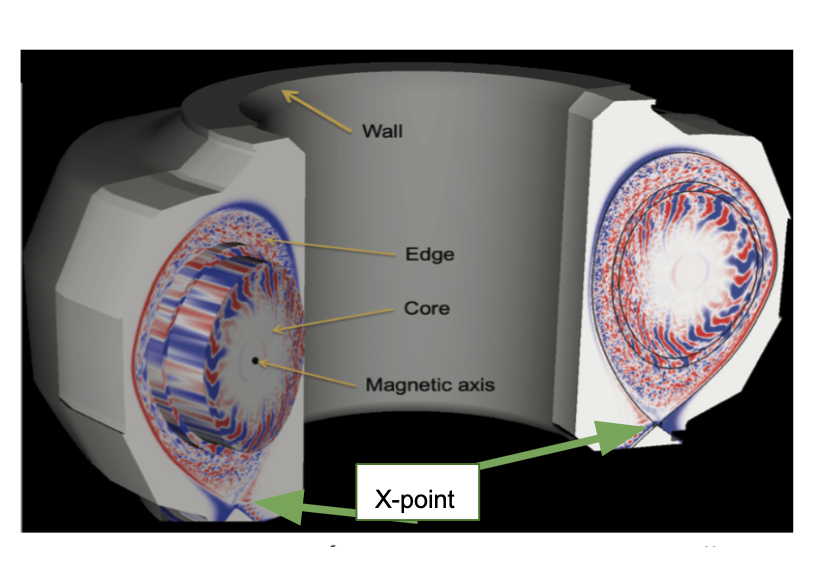} &
\includegraphics[width=0.45\textwidth]{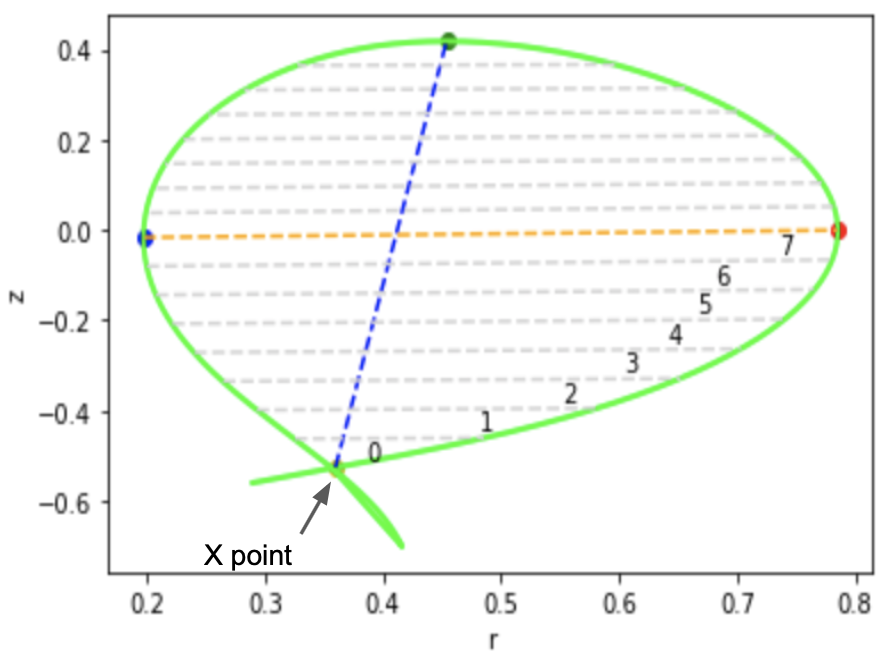}\\
 (a) & (b)
\end{tabular}
\caption{(a) Cross section of a tokamak overlaid with simulated temperature from XGC~\cite{xgc-tokamak}. (b) Outline of separatrix (green) with 8 regions in the fourth quadrant used for diffusion calculations.
}
\label{fig:outline}
\end{figure}

\begin{figure}
\centering
\begin{tabular}[b]{c}
\includegraphics[width=0.70\textwidth]{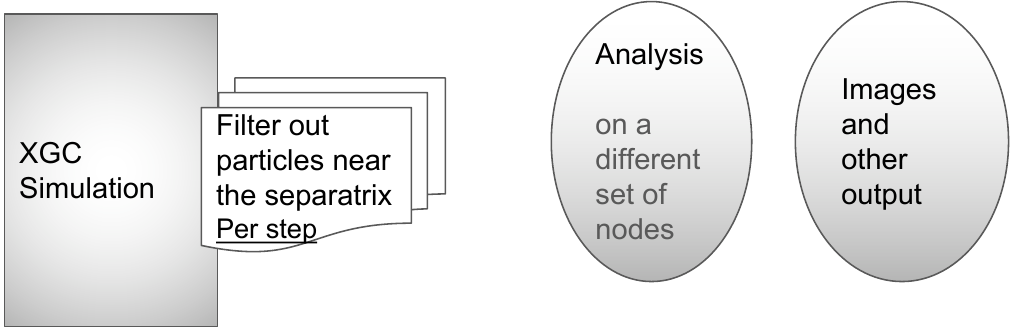} \\
(a) \\ \\
\includegraphics[width=0.70\textwidth]{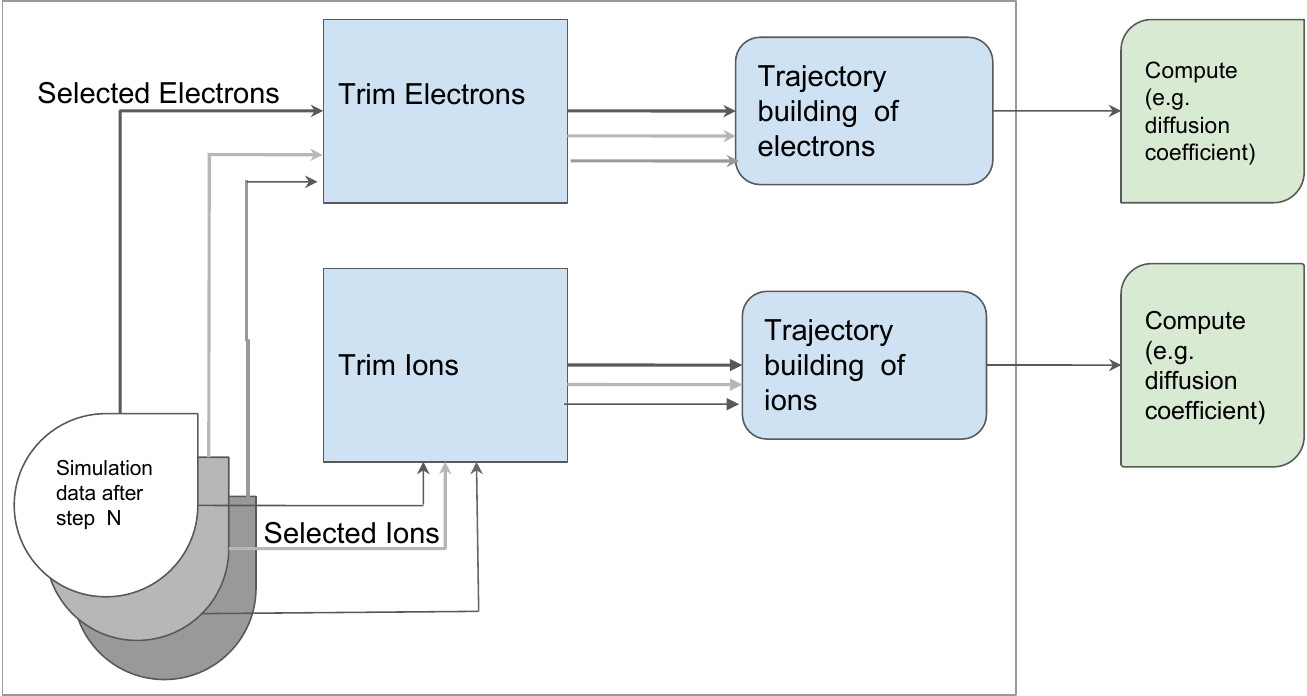}  \\
(b)
\end{tabular}
\caption{ (a) Outlines  the workflow for XGC simulation and analysis.  The arrow indicates some data is shipped for \textit{in situ} processing. (b) A detailed illustration of diffusion analysis tasks.  The large box contains the analysis tasks for step N, while simulation is working on step N+1.  Different number of nodes could be used for processing electrons and ions, in one case, we used 64 nodes for simulation, and 6 nodes for ions and 4 nodes for electrons.}
\label{fig:workflow}
\end{figure}

\paragraph{XGC}
Magnetic confinement fusion developing under large research projects such as the ITER  is expected to achieve stable controlled fusion with significant energy gain~\cite{Chang:2015:ELM,Kube2022diverter}.
The X-point Gyrokinetic Code (XGC)~\cite{xgc2008,XGC, xgc2009} is a critical simulation tool for understanding the dynamics of the plasma particles near the edge of the magnetic confinement that are essential for this fusion performance~\cite{Chang:2015:ELM}.
In particular, the particle and energy transport process near the separatrix (surface which separated the region of confined and unconfined plasma) might have significant variation, especially near the X-point (point on the separatrix where the poloidal magnetic field has a null) as shown in Fig.~\ref{fig:outline} (a).
And Fig.~\ref{fig:outline}(b) shows a simplified illustration of a cross section of a tokamak, with separatrix shown in green.
The two ends of the separatrix intersect a specially designed region of the tokamak known as the diverter.
To study the variation of the transport process (specifically diffusion) along the separatrix, we will be dividing the separatrix into different regions as illustrated in Fig.~\ref{fig:outline}(b). 

To study this diffusion process carefully, the XGC simulation will need to track massive numbers of particles~\cite{xgc-tokamak}.
Since the masses of the ions and electrons are very different, their motion is also very different, this work will track the electrons and ions separately.
Fig.~\ref{fig:workflow} shows an illustration of the overall process of our \textit{in situ} diffusion calculation, where Fig.~\ref{fig:workflow}(a) shows the overall \textit{in situ} workflow and Fig.~\ref{fig:workflow}(b) provides a high-level view of the main processing steps on the analysis nodes.
We will discuss the analysis computation in more detail in the next section.

\paragraph{ADIOS}
Our \textit{in situ} processing is carried out with the Adaptable I/O System (ADIOS), which is a high performance I/O framework with a strong support for various forms of \textit{in situ} processing~\cite{ADIOS-OLCF, softwareX, perlmutter22}.
Since the XGC code is already using ADIOS for its I/O operations, supporting additional \textit{in situ} processing does not require significant changes.

The key metaphor for describing the data during \textit{in situ} processing is the concept of ``steps,'' which matches very well with the simulation time steps in XGC.
At the end of each simulation step, information about selected particles are given to ADIOS to be passed to analysis nodes.
Depending on exactly how these processing operations are carried out, various terms are used to describe them~\cite{wolf2019scalable, 10.1111:cgf.12930, franz_smoky}.
In this work, we will use the term \textit{in situ} processing and focus on what we believe to be the distinctive features of this work: control memory usage and the quantities to be computed.

\section{In situ Particle Sorting to Build Trajectories}
\label{sec:analysis}
As described earlier, the main objective of our \textit{in situ} analysis is to support the calculation of the diffusion coefficients from important particle properties such as energy and position.
Because of the mass difference between electrons and ions in the fusion plasma, we track the diffusion of erelectrons and ions separately.
Conceptually, each simulation time step is to update the properties of plasma particles, such as position (recorded in toroidal angle, poloidal angle, minor radius (r), etc.), velocity (v), energy (E), as well as various flags such as the one used to select the particles for our diffusion calculation to be explained later.
This time-stepping behavior offers an opportunity for us to examine the state of the particles and select those that need to be tracked carefully for our diffusion study.
Next, we describe the three main steps: particle selection, trajectory building, and diffusion calculation. 

\begin{figure}
\centering
  \begin{tabular}[b]{c}
\includegraphics[width=0.75\textwidth]{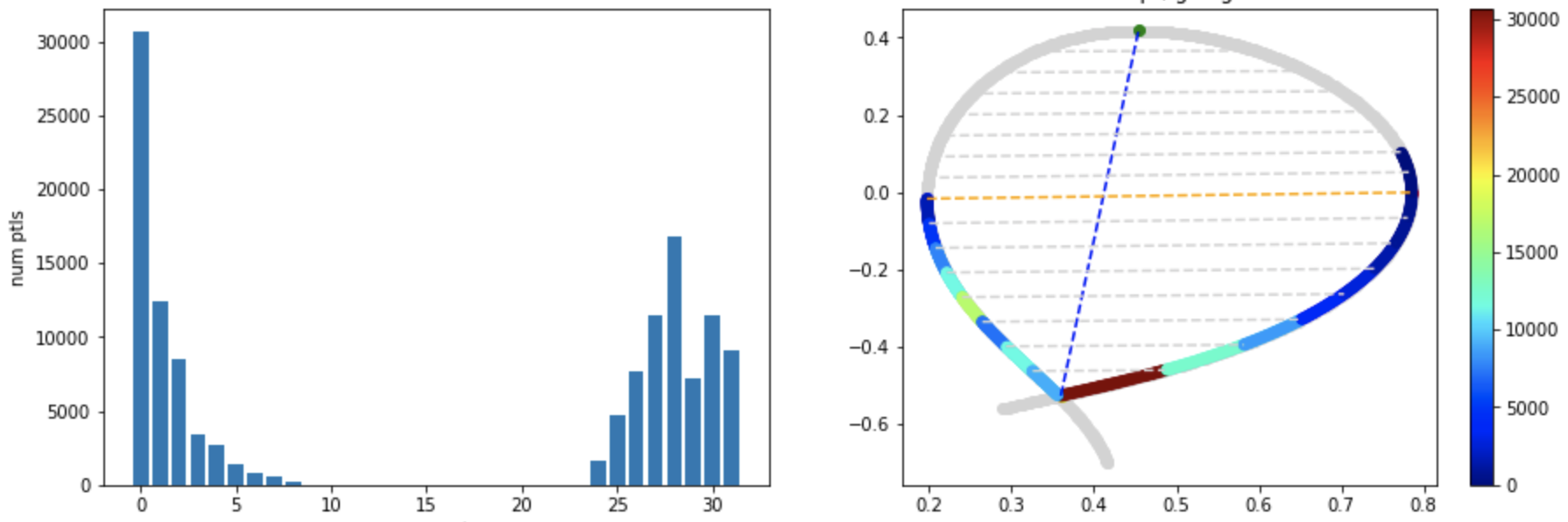}\\
(a) \\ \\
\includegraphics[width=0.75\textwidth]{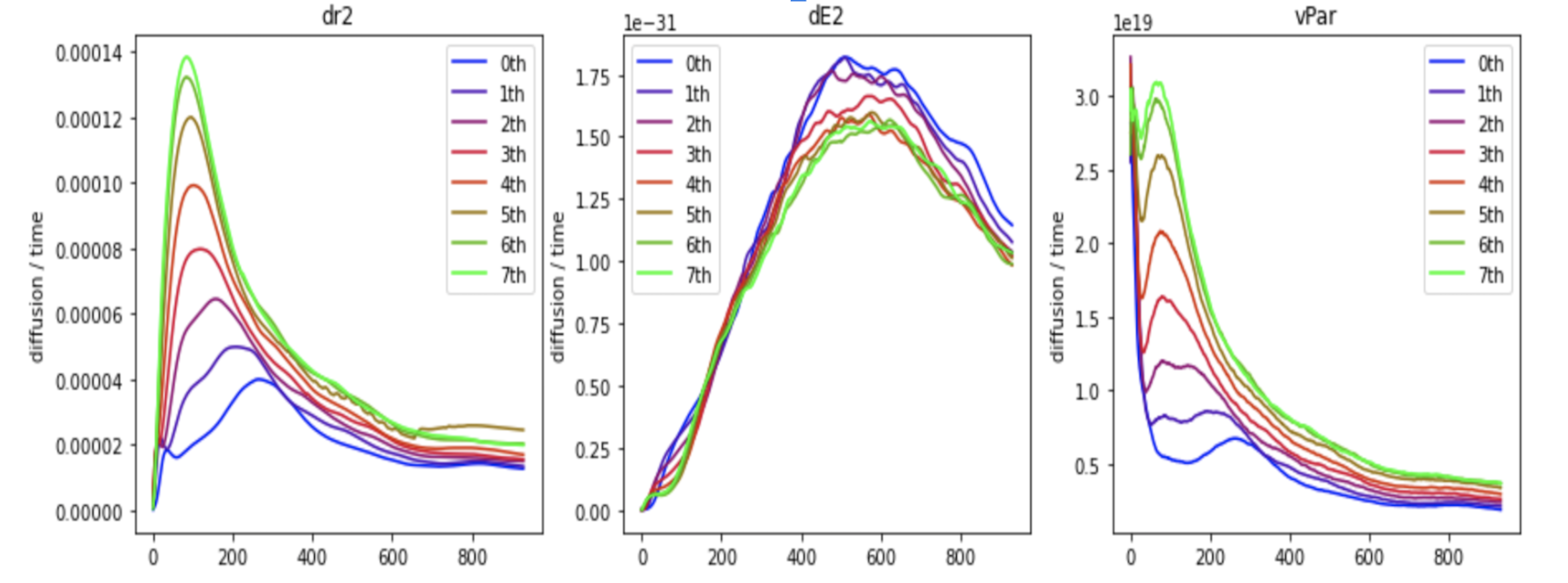} \\
(b)
\end{tabular}
\caption{Particle tracing plots.  (a) Distribution of ions in the 32 segments along the separatrix. (b) The mean square displacement over time for dr, dE, and vPar for electrons in the 4th quardrant of the separatrix.} 

\label{fig:plot_1-3}
\end{figure}

\subsection{Particle selection}
Our initial attempt was to store statistical properties, at every step,  into a 2D array for post processing.
This required extra memory usage (for the statistical data) in the simulation code, and  additional time to gather and  compute these information at the simulation nodes.
Crucially, this approach will need a post-processing step to build particle trajectories to compute the desired diffusion quantities.
This approach requires a significant increase in the memory required per particle as well as considerable time for computing the statistics and storing these quantities to files.
Overall, this was judged to an expensive option.

As a more efficient alternative, we decided to select the relevant particles and offload all processing tasks to another set of nodes as shown in Fig. \ref{fig:workflow}(a).
This approach allows the analysis nodes to perform complex calculations without slowing down the simulation.
We choose to send the relevant data through the ADIOS I/O library \cite{ADIOS-GITHUB} because this I/O library is already used by XGC~\cite{perlmutter22}.  
ADIOS uses the same API for a user code to either store data in a local file system or send data  to some (other) compute nodes.
Depending where is the data sent for analysis, the analysis process is variously termed \textit{in situ} processing, \textit{in transit} processing, and so on.
In this work, we use the term of \textit{in situ} processing to describe the distributed workflow.
As in many other \textit{in situ} processing use cases, this approach reduces the dependencies between simulation and analysis.
It is a more effective way to conduct complex analysis at high temporal resolution.

The particles of interest for our diffusion study are those with a high likelihood of escaping magnetic confinement.
The separatrix marks the boundary where the magnetic field lines inside are closed while the magnetic field lines outside are open.
The charged particles, electrons and ions, generally follow the magnetic field lines -- inside separatrix, the particles mostly follow the magnetic field lines and stay inside, while these outside of the separatrix would likely to escape the tokamak.
As such, the distance from the separatrix is a good proxy for the likelihood of a particle escaping.
The XGC code has provided a special flag for measuring this distance and provides a simple threshold on this distance measure along with additional physics-based considerations (described next) for particle selection. 

To compute the diffusion quantities effectively, we need to ensure the particle trajectories captures are computed accurately.
There are a few known scenarios where the the computed properties such as position, velocity, and so on, might not be accurate.
For example, when a particle escapes and hits the diverter plate, the computed properties such as position and velocity could include sharp discontinuities that are as unrealistic.
To deal with these special cases, the XGC developers have added a separate particle property known as the diffusion weights\footnote{Currently, there are three different weights: w0, w1, and w2, where the unwanted particles have \(w0  = -1\).}.

\begin{figure}
\centering
\begin{tabular}[b]{c}
\includegraphics[width=0.50\textwidth]{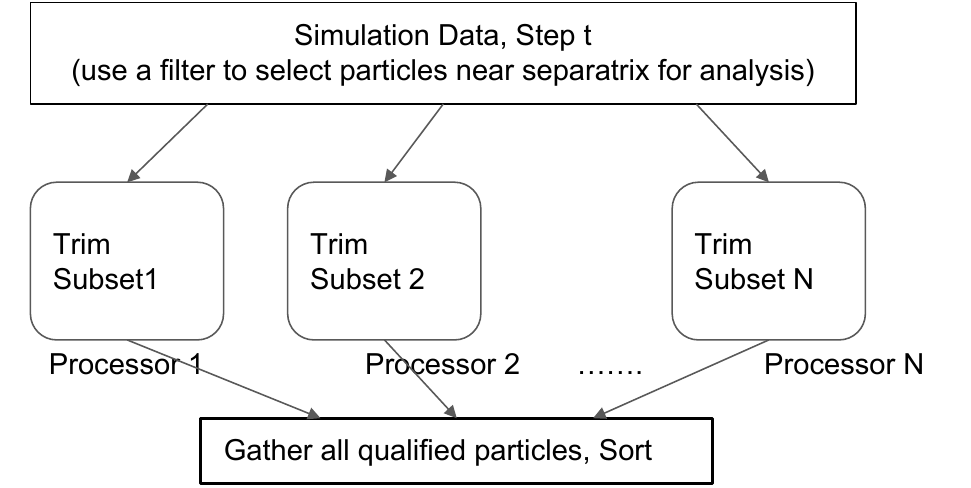}
\\
(a)\\
\\
\includegraphics[width=0.50\textwidth]{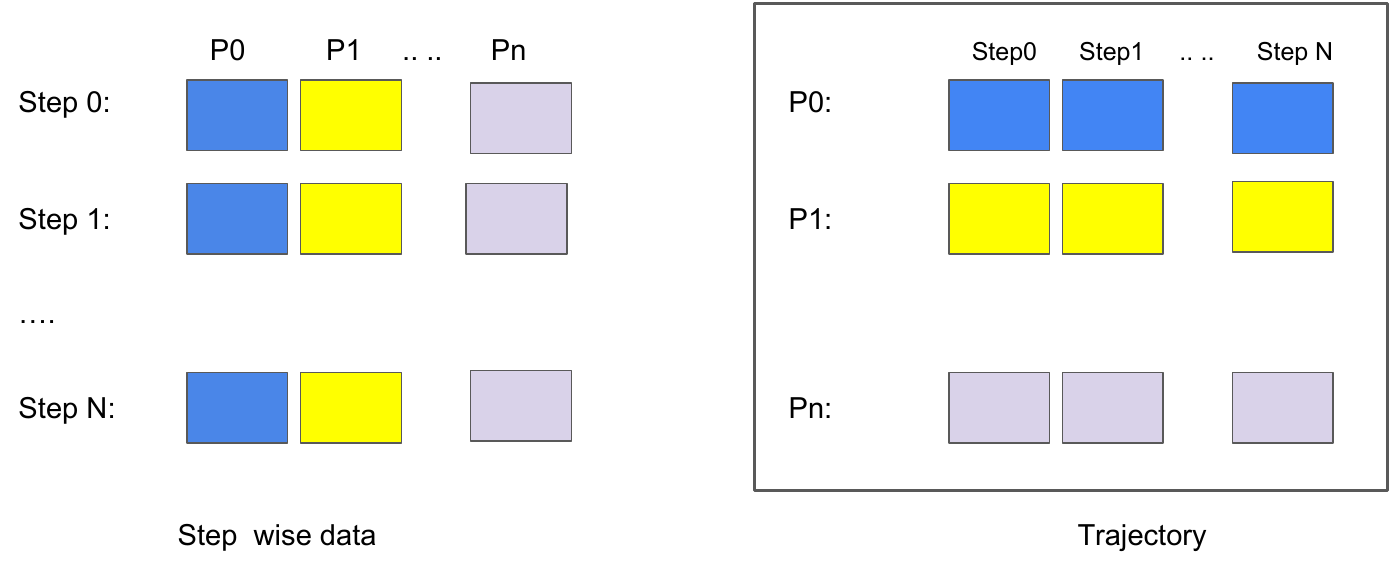}
\\
(b) \\ 
\\
\includegraphics[width=0.50\textwidth]{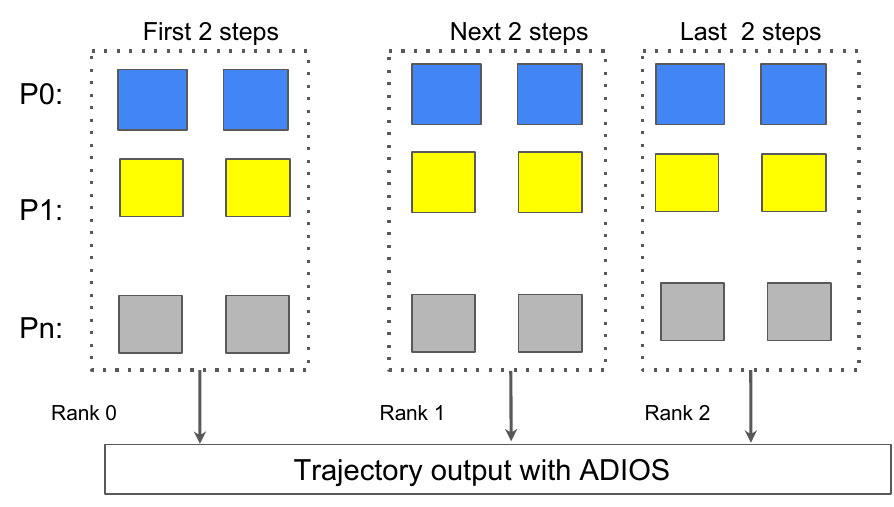}
\\
(c)
\end{tabular} 
\caption{(a) Outlines how a multi-processor analysis node removes particles from trajectory building process. 
(b) Shows how the trajectory is built from step-wise data.
(c) An illustration of how a set of trajectories with 6 steps are divided onto  3 MPI ranks.}
\label{fig:trim}
\end{figure}

\subsection{Trajectory building}
Recall that the diffusion quantities are computed from particle trajectories and the trajectory information for each particle is produced from the XGC simulation one time step at a time.
After each simulation step, the new positions and other properties become available; such information would be added to the trajectories of selected particles.
As mentioned above, XGC uses flags to reflect 
each particle's current status.
To simplify the particle tracking process, we currently keep only the trajectories whose starting positions are within a certain distance from the separatrix. 
As the simulation proceeds, more particles might drift near the separatrix, which could lead to a growing number of trajectories to compute.
At this time, we do not track these that came near the separatrix  at later time steps for our diffusion calculations. 
This implies that at the analysis nodes, upon receiving data of a new step, the first action is to trim the unwanted particles.
Fig.~\ref{fig:trim}(a) outlines how trimming works. 
Because the particle ordering is random at each step, we sort the trimmed data set by particle ID. 
At the end of the simulation, the particle trajectories can be built from  the sorted  data,  as seen in Fig. \ref{fig:trim}(b).

The above trajectory building process is also a particle sorting process, because the ultimate output could be regarded as the sorted list of particle trajectories.
In our use case,  we keep the number of particles to be tracked relatively small, even though we track the selected particles for every step of the simulation,  a global sorted list of particle IDs is still small enough to be kept on each analysis rank.
We use this global sorted list of IDs to filter out unused particles at every step to keep the memory requirement relatively low.
In the current implementation, we use rank 0 to gather and sort the particle IDs.
For larger use cases where the sorting could not be done with a single analysis node, we place to use a parallel sorting procedure known as SDS-Sort~\cite{Dong:2016:SSD}.

At the end of the simulation, the analysis nodes have step-wise properties of a set of sorted  particles.
The trajectory of each particle is built by rearranging the data layout as seen in Fig.~\ref{fig:trim} (b). 
Although  the size of the traced particle set is fixed in our use case, the data size grows as the number of time steps increases, and may not fit in a single node.
A scalable solution is to use a few MPI ranks to build the trajectories, as shown in Fig.~\ref{fig:trim}(c).
At the end of the simulation, this trajectory data set will be stored on disk to be accessed for further exploration. 

As an example, a simulation run on 64 Summit nodes (using 6 GPU per node) took 6 hours to finish 900 time steps. 
During this simulation, the number of tagged electrons increased from 33k to 28 million.  For ions, it went from 272k to 140 million.  
After trimming and sorting, the trajectory data is about 2 GB for electrons and  17 GB for ions. 
The total time on the analysis node to build the trajectory data sets is less than 2.5 hours  using 16 MPI ranks on each node.
The \textit{in situ} analysis would complete very shortly after the main simulation code.

\subsection{Diffusion calculation}
With particle information of all steps ready, we now compute the diffusion quantities.
As illustrated in Fig.~\ref{fig:outline}(b), the diffusion quantities are computed per region along the separatrix.
Let \textit{ptls} be all the particles present  in a region.
The diffusion coefficient for a property p (e.g. Energy,  velocity) is  defined below (a slightly different formula for  property  $\psi$.)   
\begin{align*}
     d(p) &= (MSQ(p) -  M(p)^2)/(\Delta t  * steps) \\ 
     d(\psi) &= (MSQ(\psi) -  M(\psi)^2)/(\Delta t  * steps * dpdrs^2) 
s\end{align*}
where $steps$ denote the number of steps taken by XGC, w0 is weight, and $dpdrs$ is the normalized $\psi$.
Furthermore, in the above equations MSQ(p) and M(p) are:
\begin{align*}
MSQ(p) =&  \sum_{ptls}(\Delta(p)^2*w0)/\sum_{ptls}(w0) \\
M(p) =& \sum_{ptls}(\Delta(p)*w0)/\sum_{ptls}(w0)
\end{align*}
The expression $MSQ(p) -  M(p)^2$ in the above equations is also known as the mean square displacement.

For our diffusion calculations, each quadrant of the separatrix is divided into 8 regions.   Fig.~\ref{fig:workflow}(b) illustrated these 8 regions in the fourth quadrant.
The actual computation procedure are programmed using python as a Jupyter notebook.
It reads in the ADIOS trajectory file, then gathers particles in  each region, computes the mean square displacement and diffusion coefficient for given properties of particles over all time step, and finally performs the visualization.
For example, Fig.~\ref{fig:plot_1-3}(a) shows the distribution of  mean square  displacement of E (energy) of electrons at time step near the end of a 64 node simulation run on the Oak Ridge Leadership Computing Facility (OLCF) Summit \cite{SummitIntro} supercomputer. 
Fig. \ref{fig:plot_1-3}(b) shows how diffusion coefficients change in regions 0-7 as time evolves.

\begin{figure}
\centering
\begin{tabular}[b]{c}
\includegraphics[width=0.8\textwidth]{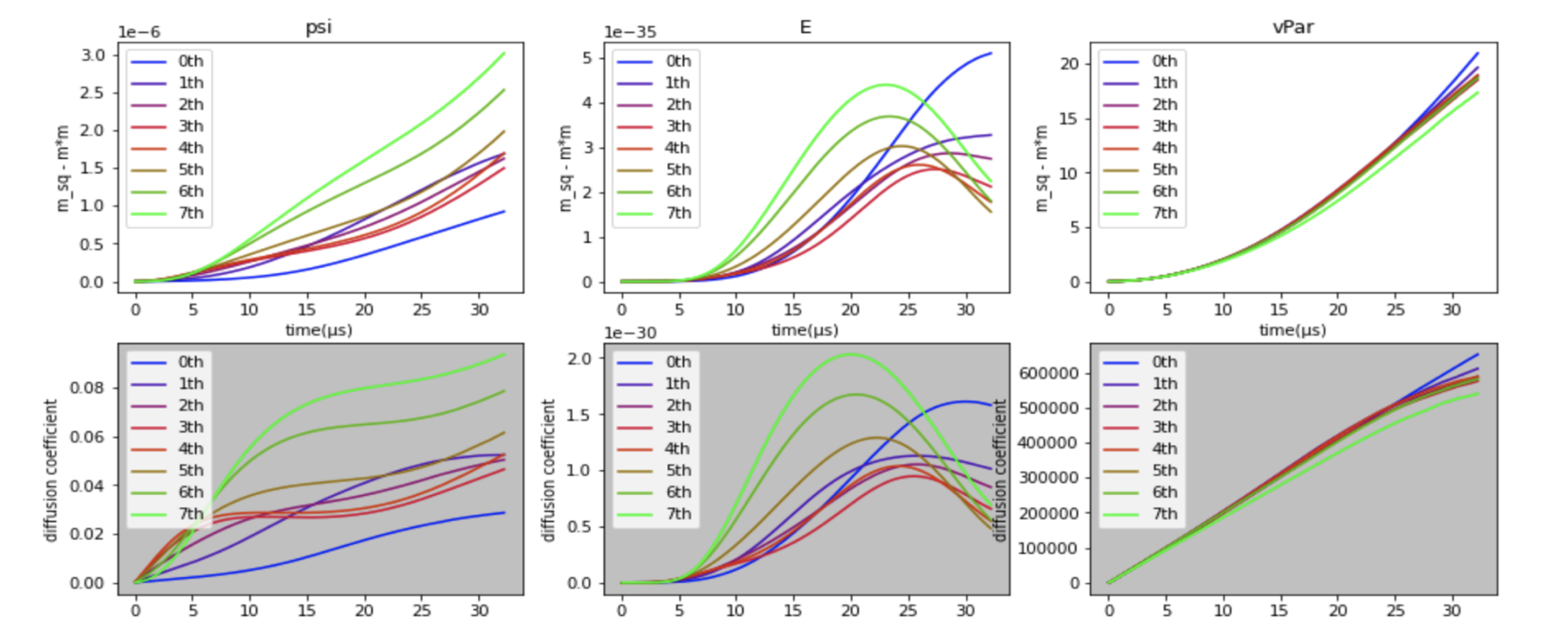}\\
(a)\\
\\
\includegraphics[width=0.8\textwidth]{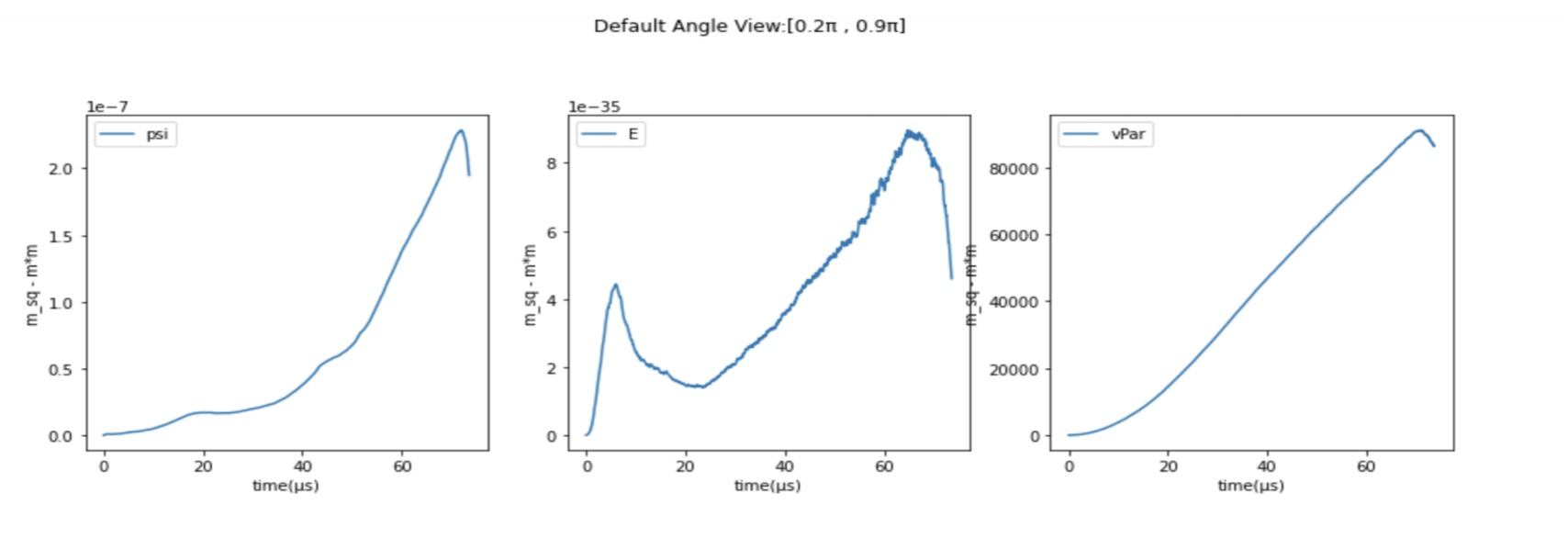}\\
(b)
\end{tabular}  
\caption{(a): Mean square displacement (top) and diffusion coefficient (bottom) are plotted  for psi ($\psi$), E and vPar for ions at regions 0-7.  (b): Poloidal angle Angle View for electrons with the default view with range from 0.2$\pi$ to 0.9$\pi$.}  
\label{fig:plot_4-5}
\end{figure}

A  region on the separatrix  could also be specified through the  poloidal angles,  where the initial location (angle 0) can either start at the X-point of the separatrix, or at the horizontal direction (which is often the default choice in most polar coordinate systems).
Through the poloidal angel, the whole separatrix has range [0, 2$\pi$].  We can compute the diffusion coefficients,  given a range of  poloidal angle.  For  example, Fig.\ref{fig:plot_4-5}(b) shows the diffusion coefficients of $\psi$, E, vPar\footnote{The variable vPar is the velocity component parallel to the magnetic field.} of the electrons with range [0.2$\pi$, 0.9$\pi$]  using default angles. 

\section{Summary and Future Work}
We have started an exploration to compute diffusion quantities for fusion plasma through an \textit{in situ} workflow.
The diffusion of hot plasma particles through the separatrix toward the diverter of a tokamak is suspected to be a cause for the collapse of fusion plasma.
Therefore, understanding the diffusion process around the separatrix is important to achieve stable magnetic confined fusion.

This work connects a state of the art simulation tool for understanding plasma particles with a novel analysis capability using the ADIOS software for data transport operations.
Our exploration indicates that the \textit{in situ} approach is able to control the memory requirement while capturing the most important particles relevant to the diffusion process.

As the collaboration with plasma physicists deepens, we anticipate the requirements on particle selection, trajectory building, and diffusion calculations would evolve and requiring the \textit{in situ} processing capability to evolve as well.
We are looking forward to these changes that might lead to a much better understanding of the particle diffusion process.

\section*{Acknowledgements}
For this research, LBNL and ORNL scientists were supported by the Exascale Computing Project (17-SC-20-SC), a joint project of the U.S. Department of Energy's Office of Science and National Nuclear Security Administration. 
PPPL scientists were supported by DOE FES and ASCR via the SciDAC project High-fidelity Boundary Plasma Simulation, under Contract No.~DE-AC02–09CH11466 to Princeton University for Princeton Plasma Physics Laboratory.
This research used INCITE resources of the Oak Ridge Leadership Computing Facility, a DOE Office of Science User Facility supported under Contract No.~DE-AC05-00OR22725, and of the National Energy Research Scientific Computing Center (NERSC), a U.S. Department of Energy Office of Science User Facility operated under Contract No.~DE-AC02-05CH11231.

\bibliographystyle{plain}
\bibliography{ref}

\begin{thebibliography}{10}

\bibitem{ADIOS-GITHUB}
Adios github page.
\newblock \url{https://github.com/ornladios/ADIOS}.

\bibitem{10.1111:cgf.12930}
Andrew~C. Bauer, Hasan Abbasi, James Ahrens, Hank Childs, Berk Geveci, Scott
  Klasky, Kenneth Moreland, Patrick O'Leary, Venkatram Vishwanath, Brad
  Whitlock, and E.~W. Bethel.
\newblock In situ methods, infrastructures, and applications on high
  performance computing platforms.
\newblock {\em Computer Graphics Forum}, 2016.

\bibitem{xgc2009}
C.-S. Chang and S.-H.~Ku et~al.
\newblock Whole-volume integrated gyrokinetic simulation of plasma tur- bulence
  in realistic diverted-tokamak geometry.
\newblock {\em In Journal of Physics: Conference Series}, 180:012057, 2009.

\bibitem{XGC}
C.~S. Chang, S.~Klasky, J.~Cummings, R.~Samtaney, A.~Shoshani, L.~Sugiyama,
  D.~Keyes, S.~Ku, G.~Park, S.~Parker, N.~Podhorszki, H.~Strauss, H.~Abbasi,
  M.~Adams, R.~Barreto, G.~Bateman, K.~Bennett, Y.~Chen, E.~D. Azevedo,
  C.~Docan, S.~Ethier, E.~Feibush, L.~Greengard, T.~Hahm, F.~Hinton, C.~Jin,
  A.~Khan, A.~Kritz, P.~Krsti, T.~Lao, W.~Lee, Z.~Lin, J.~Lofstead,
  P.~Mouallem, M.~Nagappan, A.~Pankin, M.~Parashar, M.~Pindzola, C.~Reinhold,
  D.~Schultz, K.~Schwan, D.~Silver, A.~Sim, D.~Stotler, M.~Vouk, M.~Wolf,
  H.~Weitzner, P.~Worley, Y.~Xiao, E.~Yoon, and D.~Zorin.
\newblock Toward a first-principles integrated simulation of {Tokamak} edge
  plasmas - art. no. 012042.
\newblock {\em Scidac 2008: Scientific Discovery through Advanced Computing},
  125:12042--12042, 2008.

\bibitem{xgc2008}
C.-S. Chang and S.-H. Ku.
\newblock Spontaneous rotation sources in a quiescent tokamak edge plasma.
\newblock {\em Physics of Plasmas (1994-present)}, 15:062510, 2008.

\bibitem{xgc-tokamak}
J.~Dominski, J.~Cheng, G.~Merlo, V.~Carey, R.~Hager, L.~Ricketson, J.~Choi,
  S.~Ethier, K.~Germaschewski, and S.-H.~Ku et~al.
\newblock Spatial coupling of gyrokinetic simulations, a generalized scheme
  based on first-principles.
\newblock {\em Physics of Plasmas}, 28:022301, 2021.

\bibitem{Dong:2016:SSD}
Bin Dong, Surendra Byna, and Kesheng Wu.
\newblock {SDS}-sort: Scalable dynamic skew-aware parallel sorting.
\newblock In {\em Proceedings of the 25th ACM International Symposium on
  High-Performance Parallel and Distributed Computing}, HPDC'16, pages 57--68,
  New York, NY, USA, 2016. ACM.

\bibitem{perlmutter22}
Junmin Gu, Greg Eisenhauer, Scott Klasky, Norbert Podhorszki, Ruonan Wang, and
  Kesheng Wu.
\newblock Exploring large all-flash storage system with scientific simulation.
\newblock In {\em Proceedings of the 34th International Conference on
  Scientific and Statistical Database Management}. SSDBM, 2022.

\bibitem{Chang:2015:ELM}
G.~T.~A. Huijsmans, C.~S. Chang, N.~Ferraro, L.~Sugiyama, F.~Waelbroeck, X.~Q.
  Xu, A.~Loarte, and S.~Futatani.
\newblock {Modelling of edge localised modes and edge localised mode control}.
\newblock {\em Physics of Plasmas}, 22(2):021805, 01 2015.

\bibitem{Kube2022diverter}
Ralph Kube, Michael Churchill, Seung~Hoe Ku, Jong Choi, and CS~Chang.
\newblock Plasma transport by turbulent homoclinic tangles in diverted tokamak
  plasmas.
\newblock https://meetings.aps.org/Meeting/DPP22/Session/PP11.104, 2022.

\bibitem{SummitIntro}
S.~Oral, S.S. Vazhkudai, F.~Wang, C.~Zimmer, C.~Brumgard, J.~Hanley,
  G.~Markomanolis, R.~Miller, D.~Leverman, S.~Atchley, and V.V. Larrea.
\newblock End-to-end i/o portfolio for the summit supercomputing ecosystem.
\newblock In {\em Proceedings of the International Conference for High
  Performance Computing, Networking, Storage and Analysis}, SC '19, New York,
  NY, USA, 2019. Association for Computing Machinery.

\bibitem{franz_smoky}
Franz Poeschel, Juncheng E, William~F. Godoy, Norbert Podhorszki, Scott Klasky,
  Greg Eisenhauer, Philip~E. Davis, Lipeng Wan, Ana Gainaru, and Junmin~Gu
  et~al.
\newblock Transitioning from file-based hpc workflows to streaming data
  pipelines with openpmd and adios2.
\newblock In {\em Driving Scientific and Engineering Discoveries Through the
  Integration of Experiment, Big Data, and Modeling and Simulation}, 2022.

\bibitem{ADIOS-OLCF}
The~ADIOS project.
\newblock \url{https://www.olcf.ornl.gov/center-projects/adios/}, circa 2018.
\newblock OLCF Website.

\bibitem{softwareX}
F.Godoy William, Podhorszki Norbert, Wang Ruonan, Atkins Chuck, Eisenhauer
  Greg, Gu~Junmin, Davis Philip, Choi Jong, Germaschewski Kai, Huck Kevin,
  Huebl Axel, Kim Mark, Kress James, Kurc Tahsin, Liu Qing, Logan Jeremy, Mehta
  Kshitij, Ostrouchov. George, Parashar Manish, Poeschel Franz, Pugmire David,
  Suchyta. Eric, Takahashi Keichi, Thompson Nick, Tsutsumi Seiji, Wan Lipeng,
  Wolf Matthew, Wu~Kesheng, and Klasky Scott.
\newblock {ADIOS 2}: The adaptable input output system. a framework for
  high-performance data management.
\newblock {\em SoftwareX}, 12:100561, 2020.

\bibitem{wolf2019scalable}
Matthew Wolf, Jong Choi, Greg Eisenhauer, St{\'e}phane Ethier, Kevin Huck,
  Scott Klasky, Jeremy Logan, Allen Malony, Chad Wood, Julien Dominski, et~al.
\newblock Scalable performance awareness for in situ scientific applications.
\newblock In {\em 2019 15th International Conference on eScience (eScience)},
  pages 266--276. IEEE, 2019.

\end{thebibliography}

\end{document}